\documentclass[preprint2]{aastex}

\usepackage{amsmath}
\usepackage{amssymb}
\usepackage{xspace}

\newcommand{\Msun}{{\ensuremath{\mathrm{M}_{\odot}}}\xspace}

\newcommand{\lSect}[1]{{\label{sec:#1}}}

\newcommand{\Sectff}[1]{{\ref{sec:#1}}}
\newcommand{\Sect}[1]{{\S~\Sectff{#1}}}

\newcommand{\ltaprx}{{\ensuremath{\lesssim}}\xspace}

\shorttitle{Carbon Ignition in Type Ia Supernovae}
\shortauthors{Woosley et al.}

\begin{document}

\title{Carbon Ignition in Type Ia Supernovae: An Analytic Model}

\author{S.~E.\ Woosley}

\vskip 0.2 in
\affil{Department of Astronomy and Astrophysics,
University of California, Santa Cruz, CA 95064}
\email{woosley@ucolick.org}

\vskip 0.2 in

\author{S. Wunsch}
\affil{Combustion Research Facility \\
Sandia National Laboratories, Livermore, CA 94551-0969}
\email{sewunsc@ca.sandia.gov}

\vskip 0.4 in
\and

\author{M. Kuhlen}
\vskip 0.2 in
\affil{Department of Astronomy and Astrophysics,
University of California, Santa Cruz, CA 95064}
\email{mqk@ucolick.org}

\begin{abstract}

The observable properties of a Type Ia supernova are sensitive to how
the nuclear runaway ignites in a Chandrasekhar mass white dwarf - at a
single point at its center, off-center, or at multiple points and
times. We present a simple analytic model for the runaway based upon a
combination of stellar mixing-length theory and recent advances in
understanding Rayleigh-Benard convection.  The convective flow just
prior to runaway is likely to have a strong dipolar component, though
higher multipoles may contribute appreciably at the very high Rayleigh
number (10$^{25}$) appropriate to the white dwarf core.  A likely
outcome is multi-point ignition with an exponentially increasing
number of ignition points during the few tenths of a second that it
takes the runaway to develop. The first sparks ignite approximately
150 - 200 km off center, followed by ignition at smaller
radii. Rotation may be important to break the dipole asymmetry of the
ignition and give a healthy explosion.
\end{abstract}


\section{INTRODUCTION}

Despite forty years of study (Hoyle \& Fowler 1960), the mechanism
whereby a degenerate carbon-oxygen white dwarf explodes, producing a
Type Ia supernova (SN Ia), remains poorly understood (for a recent
review see Hillebrandt \& Niemeyer 2000). Early calculations assumed
that central carbon ignition would lead to a detonation (Arnett 1969)
that would incinerate the star entirely to iron. This proved
inconsistent both with observations of features in the supernova
spectrum from intermediate mass elements and with detailed
calculations of isotopic nucleosynthesis. Nowadays it is understood
that prompt detonation does not occur because the core at ignition
time is insufficiently isothermal (Woosley 1992).

Attention in recent years has thus focused on deflagrations, subsonic
burning fronts in which pressure equilibrium is maintained across the
burning interface. Though it is controversial whether the deflagration
will later make a transition to a detonation (Niemeyer \& Woosley
1997; Khokhlov, Oran, \& Wheeler 1997; Niemeyer 1999), it is
universally assumed that the runaway begins as a deflagration (Nomoto,
Sugimoto, \& Neo 1976).

There is less certainty, though, regarding exactly how and where the
runaway is ignited. Most one-dimensional (1D) calculations, because of
their imposed symmetry, obtain ignition at the center of the
star. Many 2D studies have also assumed central ignition, largely as a
matter of convenience. However recent work (Niemeyer, Hillebrandt, \&
Woosley 1996; Reinecke, Hillebrandt, \& Niemeyer 1999, 2002ab) has
highlighted the sensitivity of the supernova outcome to precisely how
the runaway is initiated - at the center or at one or more points
off-center. These results may be summarized as showing that, in the
absence of detonation, multi-point spherically-symmetric off-center
ignition gives the most robust explosions, central single-point
ignition gives weaker ones, and single-point off-center ignition gives
such a weak explosion that it fails to unbind the white dwarf on the
first attempt.

It has been recognized for some time that the critical circumstances
affecting the ignition are determined when the convecting white dwarf
core reaches a density of $2 - 3 \times 10^9$ g cm$^{-3}$ and a
temperature of about $7 \times 10^8$ K (Nomoto et al 1984; Woosley \&
Weaver 1986). For these conditions, the time scale for the increase of
nuclear energy generation becomes comparable to the convective
turnover time, both of order 10 to 100 s.  By the time any fluid
element reaches 10$^9$ K, burnings has become quicker even
than the time it takes a sound wave to cross a pressure scale height
and, for all practical purposes, carbon burns instantly to iron. The
surface of this fluid element then becomes a ``flame'', with a well
determined speed (Timmes \& Woosley 1992) and a buoyancy given
by its density decrement, $\Delta \rho/\rho \sim 15$\%. The explosion
is born.

Woosley (1990) first suggested that the much smaller buoyancy of
convective fluid elements would still lead to appreciable radial
motion, even as the convection decoupled from the burning. Thus
ignition would occur off-center at one or more points already moving
rapidly outwards. This speculation was rendered more quantitative by
Garcia-Senz \& Woosley (1995), who, using a simple parameterized
description of ``burning floating bubbles'', estimated a typical
ignition radius $\sim$200 km. They also noted that off-center ignition
would help alleviate a chronic overproduction of neutron-rich isotopes
in Type Ia supernovae, since most of the burning would take place
farther out at lower density than in centrally ignited models.

Woosley (2001) estimated the convection speeds ($\sim$100 km s$^{-1}$)
and temperature fluctuations ($\Delta T/T \sim$0.3 - 3\%) in the
convective core at the time of runaway and, comparing them to
deflagration flame speeds and accelerations due to off-center burning,
all of which are comparable, concluded that multi-point off-center
ignition was probable.  He also pointed out that the exact number and
location of points was sensitive to small variations in
the initial conditions, thus introducing some degree of chaos in the
outcome. Since such observables as the kinetic energy, nickel mass and
peak luminosity are sensitive to how the star ignites, Type Ia
supernovae, starting from nearly identical initial conditions, will
always exhibit some irreducible diversity.

The conclusion that multi-point off-center ignition is likely was
challenged by H\"oflich \& Stein (2002). Using a 2D implicit
hydrodynamics code to follow the last few hours of the white dwarf
runaway, they found no evidence for multiple spot or strong off-center
ignition. Instead their model ignited at about 30 km, virtually at the
center, and only once. Ignition was ``induced by compressional
heat''. As we shall see, their results, though a major computational
advance, may have been influenced by attempting to model a 3D,
spherical problem while carrying only a fraction of the solid angle on
a 2D grid. The resolution and Reynolds number may also have been too
low to see multi-point ignition in a simulation that was, at best,
mildly turbulent and included only a small fraction of the fluctuation
distribution function for the temperature.

In this paper and its companion (Kuhlen, Woosley, \& Glatzmaier 2003a;
Paper II), we explore the ignition of a nuclear runaways in
Chandrasekhar mass white dwarfs using two different approaches - an
analytic model (this paper) and a 3D anelastic numerical model. In the
analytic case, we are influenced by recent developments in
understanding convection in Rayleigh-Benard experiments. These
experiments show that the qualitative character of convection may
change markedly depending upon the Rayleigh number. Even at high
Rayleigh number, a persistent ``roll'' dominates the flow pattern in
Rayleigh-Benard convection. The analogue to this in a sphere is a
dipole flow pattern. Recent numerical simulations (Paper II) suggest
that such a large scale flow is also present in (non-rotating) convective
stars. If so, it affects the mechanics of white dwarf ignition in a
major way that can only be seen by carrying the entire sphere in the
calculation.

\section{AN ANALYTIC MODEL OF THE RUNAWAY}

The final stages of the carbon runaway, wherein roughly 1.1 $\Msun$ \
of the core becomes convective, goes on for well over a century. By
the time the central temperature reaches $T_8 = T/10^8 {\rm K} = 7$
and $\rho_9 = \rho/10^9 {\rm g \ cm^{-3}}$ = 2, the typical time scale
for convection to go a pressure scale height, about 450 km, is
$\sim$10 s, and has become comparable to the nuclear time scale.

\subsection{Nuclear energy generation rate and time scale}

Nuclear energy generation during carbon ignition is given 
entirely by the (highly screened) fusion of two $^{12}$C nuclei to
form, chiefly, $^{20}$Ne and $^{24}$Mg. The approximate energy
generation rate (assuming carbon burns to a mixture of 3 parts
$^{20}$Ne and one part $^{24}$Mg;  Woosley 1986) is
\begin{equation}
\begin{split}
\dot S_{\rm nuc} \ &\approx \ 6.7 \times 10^{25} 
\ {\rm X^2(^{12}C)} \ \rho_9 \ F_{\rm sc} \ \lambda_{12,12} \\
& \ \ \ \ \ \ \ \ \ \ \ \  \ \  \ {\rm erg \ g^{-1} \ s^{-1}},
\label{snuc}
\end{split}
\end{equation}
where $\lambda_{12,12}$ is the carbon fusion reaction rate 
(Caughlan \& Fowler, 1988), $F_{\rm sc}$ is the electron screening
function, and X($^{12}$C) is the mass fraction of carbon. For a range
of temperatures, $T_8$ = 6 - 8,
\begin{equation}
\lambda_{12,12} \ \approx \ 7.6 \times 10^{-16}
\left(\frac{T_8}{7}\right)^{30}.
\end{equation}
The electron screening function (Alastuey \& Jancovici, 1978) is given
by ($\rho_9$ = 1 - 3; $T_8$ = 6 - 8)
\begin{equation}
F_{\rm sc} \ \approx \ 1100 \left(\frac{\rho_9}{2}\right)^{2.3} \ 
\left(\frac{T_8}{7}\right)^{-7},
\end{equation}
so that the energy generation rate for a composition of 50\% carbon,
50\% oxygen is
\begin{equation}
\dot S_{\rm nuc} \ \approx \ 2.8 \times 10^{13} \ \left(\frac{T_8}{7}\right)^{23}
\, \left(\frac{\rho_9}{2}\right)^{3.3} \ \ {\rm erg \ g^{-1} \ s^{-1}}.
\label{enuc}
\end{equation} 
The specific energy available is
\begin{equation}
q_{\rm nuc} \ = \ 4.0 \times 10^{17} \ {\rm X(^{12}C)} \ {\rm erg \ g^{-1}}.
\end{equation}
The specific heat at constant pressure, which is required to estimate
the nuclear time scale, is (e.g., Chiu 1968)
\begin{equation}
\begin{split}
c_{\rm P} \ &= \ \left( \frac{\partial \ \epsilon}{\partial T}
                    \right)_{\rm ions} +\left( \frac{\partial \
                    \epsilon}{\partial T} \right)_{\rm electrons} +
                    \left(\frac{\partial \ \epsilon}{\partial T}
                    \right)_{\rm radiation} \\ 
&= \ \left(\frac{3 \ N_{\rm A} k}{2\bar A}\right) \ + \ \frac{\pi^2
                    k^2}{x m_e c^2} \rho N_{\rm A} Y_e \, T \ +
                    \left(\frac{4 a T^3}{\rho}\right)\\
&= \ 9.1 \times 10^{14} \ + \ \frac{8.6 \times 10^{13} \
                    T_8}{\rho_9^{1/3}} \ \\
& \ \ \ \ \ \ \ \ \ \ + \ \frac{3.0 \times 10^{9} \ T_8^3}{\rho_9} \ \
                    {\rm erg \ g^{-1} \ (10^8 K)^{-1}},
\label{heatcap}
\end{split}
\end{equation}
where $\bar A$ is the mean atomic mass number (13.7 for 50\% carbon,
50\% oxygen by mass); $m$ is the mass of the electron, $x = p_{\rm
F}/mc$ is related to the mass density by $9.74 \times 10^5 \mu_e x^3 =
\rho$; $Y_e$ is the electron mole number, here 0.5; and the other
symbols have their usual meanings. For the conditions of interest,
e.g., $T_8$ = 7 and $\rho_9$ = 2, the heat capacity of the radiation
field is negligible and the ions and electrons together provide $1.4
\times 10^{15}$ erg g$^{-1}$ (10$^8$ K)$^{-1}$. This relatively small
heat capacity makes the carbon highly incendiary in the sense that a
small amount of burning raises the temperature considerably. For lower
densities, however, $\rho_9 \sim 0.01$, the heat capacity of the
radiation field becomes important, and this is what finally keeps the
star from burning entirely to iron.

For central temperatures near $T_{o,8} = 7$, The nuclear time under
these conditions is
\begin{equation}
\begin{split}
\tau_{\rm nuc} \ &= \ \left(\frac{1}{\dot S_{\rm nuc}} \, \frac {d \,
  \dot S_{\rm nuc}}{d \, t} \right)^{-1}\\
&\approx \ \left(\frac{1}{\dot S_{\rm nuc}} \, \frac{\partial \dot S_{\rm
  nuc}}{\partial T} \, \frac{\partial T}{\partial t}\right)^{-1}\\
&= \ \frac{c_{\rm P} \, T}{23 \, \dot S_{\rm nuc}}\\
&\approx \ 15 \, (\frac{7}{T_8})^{22} \
(\frac{2}{\rho_9})^{3.3} \ {\rm s}.
\label{taunuc}
\end{split}
\end{equation}

This is the time for the energy generation in an isolated region to
increase from its starting value at temperature, $T_8$, to such high
values that the reactions are virtually instantaneous. While
convection remains efficient, this time is lengthened in the star by a
factor of approximately 50 (\Sect{pdf}).

\subsection{Luminosity}

The long convective episode in the pre-explosive star establishes an
overall adiabatic temperature gradient in the central regions. Using
this condition, the known density structure, energy generation,
Eq.\eqref{enuc}, and assuming hydrostatic equilibrium, one can
estimate the luminosity.

Because of the extreme sensitivity of the energy generation to
temperature, the luminosity will originate from a small fraction of
the mass justifying a first order polytropic extrapolation ($n$ = 3) of
the central conditions. The equation of state is that of a
relativistically degenerate gas,
\begin{equation}
\begin{split}
P \ &= \ K \, \rho^{4/3}\\ & = 1.24 \times 10^{27} \
\left(\frac{\rho_9}{2}\right)^{4/3} {\rm dyne \ cm^{-2}},
\end{split}
\end{equation}
and  the polytropic radius parameter, 
\begin{equation}
\begin{split}
a  \  &= \ \left(\frac{K}{\pi \, G \, \rho_o^{2/3}}\right)^{1/2}\\
&= \ 385 \ {\rm km} \ \left(\frac{2}{\rho_{o,9}}\right)^{1/3},
\end{split}
\end{equation}
with $\rho_{o,9}$ the central density in 10$^9$ g cm$^{-3}$.  Defining
a dimensionless radius, $\zeta = r/a$, the mass interior to radius $r$
is given by ($\zeta^2 \ll 1$),
\begin{equation}
M(r) \ \approx \ \frac{4 \pi}{3}  r^3  \rho_o \, (1 \ - \ \frac{3}{10} \, \zeta^2),
\end{equation}
and the density at radius $r$ is 
\begin{equation}
\rho(r) \ \approx  \ \rho_o \, (1 \  - \ \frac{1}{2} \, \zeta^2).
\end{equation}
Combining the equation of hydrostatic equilibrium
\begin{equation}
\frac{dP}{dr} \ = \ -\frac{G M(r)\rho(r)}{r^2},
\end{equation}
with the condition for an adiabatic temperature gradient
\begin{equation}
\frac{dT}{dr} \ = \ (1 \ - \ 1/\Gamma_2) \, (T/P)\, \frac{dP}{dr},
\end{equation}
one obtains, for $\Gamma_2 \approx 1.7$, the variation of temperature
with radius near the center of the white dwarf (Woosley 1990)
\begin{equation}
\begin{split}
T(r) \ &\approx \ T_o \, \left( 1 \ - \ \frac{\Gamma_2 -1}{\Gamma_2} \
\frac{2 \pi G \rho_o^2 r^2 f_1}{3 P_o}\right) \\
&\approx \ T_o \ \left(1 \ - \ 0.0185 \left(\frac{\rho_{o,9}}
{2}\right)^{2/3} f_1 \, r_7^2\right),
\label{tadib}
\end{split}
\end{equation}
where $f_1 \approx (1 - \frac{1}{15} \zeta^2)$ is a correction, near unity, for
the density gradient.  This equation also describes the temperature
evolution of any adiabatically expanding (or contracting) fluid
element as it moves in a region near the star's center.

One may then integrate Eq.\eqref{tadib} to obtain the luminosity as a
function of central temperature,
\begin{equation}
\begin{split}
L \ &= \ 4 \pi \, \int \, \dot S_{\rm nuc} \, \rho \, r^2 \, dr \\
&\approx \ 7.0 \times 10^{44} \ {\rm erg \ s^{-1}} \
(\frac{\rho_{o,9}}{2})^{4.3} \, (\frac{T_{o,8}}{7})^{23} \, I,
\label{loft}
\end{split}
\end{equation}
where $I$ is the integral,
\begin{equation}
I = \int r_7^2 \,(1 \ - \ b \, r_7^2)^{23} f_2 \ dr_7,
\end{equation}
with $b = 0.0185 (\rho_9/2)^{2/3} f_1$ and $f_2 \approx (1 - \frac{1}{2}
\zeta^2)^{4.3}$. This integral can be evaluated numerically to give $I
= $ 0.98 and 0.65 for $\rho_{o,9}$ = 2 and 3 respectively. The
result for the luminosity is valid to better than 20\%.

One can also estimate the size of the energy generating region by
calculating the radius where $L$ reaches one-half its value, 140 and
120 km respectively for $\rho_9$ = 2 and 3. That is, approximately
one-half of the luminosity of the star is generated in its inner 130
km.

\subsection{Results from mixing-length theory}

As the runaway proceeds, the central temperature rises and, along with
it, the luminosity. Energy is transported by convection and dissipated
by expansion, i.e., $PdV$ work against gravity, and neutrinos, both of
which occur chiefly outside the energy-generating central core.

Conditions near the end of this ramp up can be estimated using
mixing-length theory (e.g., Clayton 1983; Lantz \& Fan 1999). In
particular, the heat flux at radius $r$ is given by
\begin{equation}
\begin{split}
\phi(r) \ &= \ \frac{L(r)}{4 \pi \, r^2} \\
&= \ \frac{\rho}{2} \, v_{\rm rms} \, c_{\rm P} \, (l \Delta \nabla T) \\
&\approx   \ \frac{\rho}{2} \ v_{\rm rms} \, c_{\rm P} \, (\Delta T)
\label{flux}
\end{split}
\end{equation}
where $L(r)$ is the luminosity at radius $r$, $\rho$ is the density,
$v_{\rm rms}$, the average convective velocity there, $c_{\rm P}$, the heat
capacity, Eq.\eqref{heatcap}, $l$, a characteristic size for
the convection region, and $\Delta T$, the temperature change
across this region in excess of the adiabatic value. The factor
``$\frac{1}{2}$'' accounts for the fact that heat is carried by the
outward moving fluid elements, while lower entropy elements return the
mass. Density is assumed nearly constant in all elements at a given
radius.

Fluid elements are buoyant because, at constant pressure, their excess
temperature is accompanied by a deficiency in density.  The
logarithmic derivative of density with respect to temperature, at
constant pressure, is given by
\begin{equation}
\begin{split}
\delta_{\rm P} \ &= \ - \left(\frac {\partial \, ln \, \rho}{\partial \, ln
\, T}\right) \\
&= \ \frac{T}{\rho} \ \frac{\Delta \rho}{\Delta T}\\
&= \ \frac{T}{\rho} \ \left(\frac{\partial \, {\rm P}}{\partial T}\right)_{\rho} 
\left(\frac{\partial \, {\rm P}} {\partial \rho}\right)_T^{-1}\\
&\approx \ 1.9 \times 10^{-2} \ \frac{T}{\rho}
\label{dp}
\end{split}
\end{equation}
for a range of temperatures near $T_8 = 7$ and $\rho_9 = 2$.  The
derivatives in the above equation were evaluated numerically using the
equation of state in the Kepler code (Weaver, Zimmerman, \& Woosley
1978). For example, for $\rho_9 = 2$ and at $T_8 = 7$ and 8
respectively, $\delta_{\rm P}$ = $6.6 \times 10^{-3}$ and $7.7 \times
10^{-3}$ respectively. The small value of this dimensionless constant
reflects the extreme degeneracy of the gas, i.e., that a large
temperature change is required to give a pressure change comparable to
that resulting from a small change in density.

A typical convective velocity is then given by
\begin{equation}
v_{\rm rms} \ \approx \ \left(\frac{2 \, g \, \Delta
\rho}{\rho}\right)^{1/2} \ l^{1/2},
\label{vbar}
\end{equation}
with $g$, the local acceleration due to gravity, i.e., $G \, M(r)/r^2$,
and $\Delta \rho$, the density variation corresponding to $\Delta
T$. Combining Eqs. \eqref{flux}, \eqref{dp}, and \eqref{vbar}, one has
the mixing-length-theory estimate for the typical convection speed,
\begin{equation}
\begin{split}
v_{\rm rms} \ &\approx \ \left(\frac{4 \, g \, r \, \delta_{\rm P} \,
\phi(r)}{\rho \, c_{\rm P} \, T}\right)^{1/3} \\ &\approx \
\left(\frac{4 \, G \, \delta_{\rm P} \, L}{3 \, c_{\rm P} \,
T}\right)^{1/3}.
\label{speed}
\end{split}
\end{equation}
Here $L$ is the luminosity in erg s$^{-1}$.  So long as
$L$ is nearly constant, the result is insensitive to $l$.

For $L_{45} \ = \ L(200 \ {\rm km})/10^{45}$ erg s$^{-1}$,
\begin{equation}
v_{\rm rms} \ \approx \ 40 \ {\rm km \ s^{-1}} \ (\frac{7}{T_{o,8}})^{1/3} \,
L_{45}^{1/3}
\end{equation}
Since $L$ is zero at the stellar center, $v_{\rm rms}$ formally goes to
zero there, but its actual value depends upon the pattern of fluid
flow.  As is discussed in \Sect{flowp}, the velocity just outside the
energy generating core, say at 100 - 200 km, may be a characteristic
of the entire convection region.

Most of the temperature dependence of $v_{\rm rms}$ is contained in the
$L/T$ term. This depends on the central conditions roughly as $T_o^{22}
\rho_o^{4.3}$, so the velocity scales with the central temperature
roughly as $T_o^7 \rho_o^{1.4}$. From Eq.\eqref{loft} the luminosity for
$T_{o,8} = 7, \rho_9 = 2$ is about $10^{45}$ erg s$^{-1}$,
so
\begin{equation}
v_{\rm rms} \ \approx \ 40 \ {\rm km \ s^{-1}} \
(\frac{\rho_9}{2})^{1.4}(\frac{T_{o,8}}{7})^7.
\end{equation}
That is, one expects speeds about 2.5 times faster, or 100 km s$^{-1}$,
at $T_{o,8} = 8$ compared with $T_{o,8} = 7$.

The corresponding characteristic temperature excess across the
convection zone is
\begin{equation}
\Delta T \ = \ \left(\frac{2 \, \phi^2 \, T}{\rho^2 \, c_{\rm P}^2 \, g
\, l \, \delta_{\rm P}}\right)^{1/3},
\label{deltat}
\end{equation}
or equivalently,
\begin{equation}
\begin{split}
\frac{\Delta T}{T} \ &= \ \frac{v_{\rm rms}^2}{2\, g \, l \,
\delta_{\rm P}}\\
&= \ 0.005 \left(\frac{\rho_9}{2}\right)^{2.1} \,
\left(\frac{T_{o,8}}{7}\right)^{14}.
\label{tfluc} 
\end{split} 
\end{equation}
This is the change in temperature {\sl excess} across the
region, beyond the adiabatic value. It is much smaller than the
total change in temperature across the convection zone.

The typical density variation, $\Delta \rho/\rho$, is $\delta_{\rm P}$
times this, or about $3.6 \times 10^{-5}$.  Because of the uncertain
choice of $l$ and the use of a single energy characterized by $\Delta
T$ to carry the flow, these estimates are probably only accurate to a
factor of two. There may also be some weak dependence of $\Delta T$ on
the Rayleigh number not included in our simple analysis. 

As we shall see later, it is a characteristic of the higher
temperature fluctuations that ignite the explosion that they move with
a higher speed than the average, $v_{\rm rms}$. Also at $T_{o,8} = 7.5 -
8.0$ which may be a more typical choice for the mean central
temperature at ignition, $v_{\rm rms} \sim 100$ km s$^{-1}$. It is
interesting, and seemingly an accident of nature, that, to about a
factor of two, this is the same as the conductive laminar flame speed
(Timmes \& Woosley 1992),
\begin{equation}
u_{\rm lam} \ \approx \ 76 \ {\rm km \ s^{-1}} \
\left(\frac{\rho_9}{2}\right)^{0.805} \ \left(\frac{\rm
X(^{12}C)}{0.5}\right)^{0.889}.
\label{vlam}
\end{equation}
Partly because of this coincidence there is a dependence of the
outcome on the carbon mass fraction in the white dwarf interior.

\subsection{Some characteristic measures}
\lSect{numbers}

The opacity, given by electron conduction (Timmes 2000 and 2002,
private communication), is
\begin{equation}
\kappa_{\rm cond} \ \approx \ 2.7 \times 10^{-5} {\rm cm^2 \ g^{-1}}
\ \left(\frac{2}{\rho_9}\right)^{1.4} \ \left(\frac{T_8}{7}\right)^{2.2}.
\label{opac}
\end{equation}
Using this to get the conductivity, 
\begin{equation}
\sigma \ = \ \frac{4 \, a \, c \, T^3}{\rho \, \kappa},
\end{equation}
for typical conditions, $\sigma \approx 3 \times 10^{18}$ erg
cm$^{-1}$ K$^{-1}$ s$^{-1}$. The viscosity, $\eta$, is given (Nandkumar
\& Pethick 1984) by
\begin{equation}
\eta \ \approx \ \frac{1.9 \times 10^9}{Z} \
\left(\frac{\rho_9}{2}\right) \, I_2^{-1} \ {\rm g \ cm^{-1} \ s^{-1}},
\end{equation}
where $I_2 \approx 0.5$ and $Z \approx 7$. Hence $\eta \approx 10^9$ g
cm$^{-1}$ s$^{-1}$. From these one can calculate a Rayleigh number
(the so called ``dimensionless temperature gradient'' in convection;
the ratio of buoyancy forces to diffusion forces)
\begin{equation}
\begin{split}
Ra \ &= \ \frac{g \, l^3 \, \rho^2 \, c_{\rm P} \,\delta_{\rm P} \,
\Delta T}{T \, \eta \, \sigma} \\ &\sim \ 10^{25},
\label{rayleigh}
\end{split}
\end{equation}
where we have assumed that the appropriate $\Delta T$ is approximately
$(\Delta \nabla T) \, l$, Eq. \eqref{flux}, from mixing length theory.
One can also estimate a Prandtl number (ratio of momentum transport to
heat conduction)
\begin{equation}
\begin{split}
Pr \ &= \ \frac{c_{\rm P} \, \eta}{\sigma} \\
&\sim 4 \times 10^{-3},
\label{prandtl}
\end{split}
\end{equation}
a Reynolds number (ratio of inertial forces to viscous forces),
\begin{equation}
\begin{split}
Re \ &= \ \frac{\rho \, v_{\rm rms} \, l}{\eta} \\
&\approx \ 10^{14}, 
\label{reynolds}
\end{split}
\end{equation}
and a Kolmogorov length (where turbulence dissipates),
\begin{equation}
\begin{split}
L_{\rm Kol} \ &= \ l \ Re^{-3/4} \\
&\approx \ 3 \times 10^{-4} \ {\rm cm}. 
\label{kolmog}
\end{split}
\end{equation}
This combination of Rayleigh and Prandtl
numbers is well beyond the limits of what can be studied on the Earth
by either experiment or simulation.

One can also estimate the Nusselt number (total rate of heat
transfer compared with conduction),
\begin{equation}
\begin{split}
Nu \ &= \ \frac{\phi \, l}{\sigma \, \Delta T} \\
&\approx \ \frac{\rho \, v_{\rm rms} \, l \, c_{\rm P}}{2 \, \sigma}\\
&\approx 3 \times 10^{11}
\label{nusselt}
\end{split}
\end{equation}
which will be relevant in later discussions.

\section{LESSONS FROM RAYLEIGH-BENARD CONVECTION}
\lSect{rb}

\subsection{Recent Developments}

There exists a rich literature of experiments and simulations that
study the convection of matter between a hot and a cold plate. An
important issue is how the relation between heat transport and the
Rayleigh number, called $Nu-Ra$ scaling, itself varies with $Ra$
(e.g., Castaing et al. 1989; Grossman \& Lohse 2000; Kadanoff
2001). Another important issue is how the form of the probability
distribution function for temperature fluctuations (PDF) scales with
$Ra$ and $Pr$. Is it exponential or Gaussian?

Below about $Ra \sim 10^8$, it is thought that flow in a
Rayleigh-Benard cell has not become completely chaotic, but above
10$^8$ the regime of so called ``hard turbulence'' is encountered.
Characteristics of hard turbulence include (Xia \& Qiu 1997): 1) A Nusselt
number ($Nu$) that scales as $Ra^{2/7}$ (Castaing et al. 1989) instead
of the classically expected $Ra^{1/3}$; 2) a PDF that is exponential,
not Gaussian as might have been expected based on the central limit
theorem (Kolmogorov 1962); and 3) coherent large-scale circulation.
Experiments by Niemela et al. (2000) confirm these properties of hard
turbulence across a large range of Rayleigh numbers, 10$^7$ -
10$^{17}$. Numerical calculations by Rogers, Glatzmaier, \& Woosley
(2003) confirm that at least the $Ra$-$Nu$ scaling characteristics of
hard turbulence persist when the flow is compressible with density
stratification.

However, our white dwarf problem is characterized by $Ra$ still 10$^8$
larger than studied by Niemela et al. More importantly perhaps, we are
interested in convection without hard boundaries. Energy is dissipated
in the star by expansion and neutrinos, both volumetric losses, not
conductivity to a cold boundary. Heating also occurs over an extended
region making for a very thick ``wall zone'' and a range of
temperature fluctuations that is not limited by the temperature of
some hot plate. In the Ia, the flow can also pass right through the
core of the burning region and out the other side, allowing more
complete mixing between the core and the convective region than is
possible in Rayleigh-Benard.

A hint of what may lie ahead at {\sl very} high values of Rayleigh
number, comes from studies by Kraichnan (1962), and is referred to in
the literature as the ``ultimate'' or ``Kraichnan'' regime of
convection (Grossman \& Lohse 2000; Kadanoff 2001) . In situations
where the effects of walls are suppressed, this transition may occur
at much more modest values of $Ra$, even 10$^6$ (Lohse \& Toschi
2003). In this regime, it is thought that $Nu \propto (Ra Pr)^{1/2}$
and that the large scale ``rolls'' seen in Rayleigh-Benard convection
at lower $Ra$ may give way to a more fragmented, chaotic pattern.  It
is unknown whether the distribution of temperature fluctuations in
this regime is Gaussian or exponential.

It is noteworthy, however, that the $Nu-Ra$ scaling in this ultimate
regime recovers the same simple scaling for $\Delta T$ found from
mixing length, i.e., Eq.\eqref{deltat}. That is, Eqs.\eqref{rayleigh},
\eqref{prandtl}, and \eqref{nusselt} plus the condition
\begin{equation}
Nu \ \sim \ Ra^{1/2} Pr^{1/2}
\end{equation}
implies that 
\begin{equation}
\Delta T \ = \ \left(\frac{\phi^2 \, T}{\rho^2 \, c_{\rm P}^2 \, g
\, l \, \delta_{\rm P}}\right)^{1/3},
\end{equation}
which to a factor of order unity is Eq.\eqref{deltat}.  This
correspondence does not exist for any other scalings between $Nu$ and
$Ra$, which in general would leave some residual dependence of $\Delta
T$ on the conductivity or viscosity. The mixing length approximation
is, apparently, equivalent to Kraichnan scaling. Might the other
properties of near infinite {\sl Ra} number circulation in
Rayleigh-Benard cells also be relevant?

\subsection{Flow Patterns}
\lSect{flowp}

To a large degree, the velocities, temperature fluctuations, and,
ultimately, the ignition process depend upon the circulating pattern
assumed by the major flows within the core. We consider two
representative cases - the ``isotropic model'' and the ``dipole
model''. The actual solution may have aspects of both.

In the isotropic model, there is no preferred direction. Matter enters
the burning region from all angles, is heated, reverses its direction
and flows out. In the perfectly symmetric model, matter at the center
is at rest, but this ideal state is never realized.  Small imbalances
in flow will result in overshooting, first in one direction, then
another, so that the velocity vector at the origin varies in random
way, being zero only on the average. A slice at constant radius of the
convective flow might resemble the solar photosphere, except that the
large entropy contrast between upward moving plumes and downflows
would be absent. For high $Ra$, the characteristic radial speed near
the center would be less than, but perhaps not much less than that say
a hundred km out. In this sense, the flow pattern might resemble the
radial equivalent of Rayleigh-Benard in the ``ultimate'' or Kraichnan
regime. Though not yet studied in the laboratory because of its
inaccessibly high $Ra$, it is hypothesized that in this regime plumes
would travel from one plate to another in near ballistic fashion with
the large scale circulation suppressed (Kadanoff 2001).

One might expect that in the absence of rotation there can be no
preferred orientation, that is, this isotropic model would be the only
physical one. However, numerical experiments (Kuhlen, Woosley, \&
Glatzmaier 2003b; Paper 2) show that the convective flow in either a
sphere or thick shell often takes on a dipole character. Matter flows
in from one side, is heated in the central region, and flows out the
other, like a jet engine. The center, far from being a point of
stagnation, is characterized by the same high velocities found farther
out in the convection zone. This is the stellar analogue of the large
scale circulation, or ``rolls'', seen in Rayleigh-Benard convection.
Similar dipole flows have been seen in three-dimensional studies of red
giant convection by Woodward and colleagues (Porter, Anderson, \&
Woodward 1997; Porter, Woodward, \& Jacobs 2000; Woodward, Porter, \&
Jacobs 2002, 2003).

This dipole circulation is an example of spontaneous symmetry
breaking. In the absence of rotation, the dipole picks an arbitrary
axis, determined ultimately by tiny perturbations in the initial
model, and maintains it for many convective turnover times. Over very
long periods, the orientation of the dipole may vary due to the
occasional large ``intermittent'' occurrence. Also, the average over
many calculations with random starting conditions would give no
preferred angle, so, in a sense, the model is still isotropic. But
during the time the runaway develops, the main flow may be highly
directional and this has important implications for the explosion that
follows.

\section{IGNITION}
\lSect{ign}

The runaway first commences when the temperature of the hottest
fluctuation, $T_{\rm max}$, leads to nuclear heating faster than the
adiabatic cooling that occurs when a blob crosses the burning region.
Ignition will thus occur when the integral along a convective path,
\begin{equation}
\int \ \left[\left(\frac{d \, T}{d \, r}\right)_{\rm
exp} \ + \ \frac{\dot S_{\rm nuc}}{c_{\rm P} \, v_{\rm rms}} \,
\right] \ dr,
\end{equation}
diverges. From Eq. \eqref{tadib}
\begin{equation}
\left(\frac{d \, T}{d \, r}\right)_{\rm exp} \ \approx \ - 0.037 
\, T_c \, (\frac{\rho_9}{2})^{2/3} \, r_7.
\label{gradt}
\end{equation}

It remains to specify the velocity and some distribution of starting
temperatures at the center of the star. As we shall see shortly
(\Sect{multip}), the average central temperature at ignition is in the
range $T_{o,8} = 7.5 - 8.0$. From Eq. \eqref{vbar}, $v_{\rm rms} \approx 60
- 100$ km s$^{-1}$. Taking 80 km s$^{-1}$ as representative, we find
that the first runaway will occur at a radius of over 300 km for a
fluctuation hat started with a central temperature $T_{\rm max,8} =
8.623...$.

This estimate is a little large.  The initial perturbations in the
star's center cannot have their temperature specified to arbitrary
accuracy. As we shall see in \Sect{multip}, the typical variation in
the temperature of the hottest few points in the core is a fraction of
$\Delta T$, Eq.\eqref{deltat}, about 1\%. Thus instead of $T_{\rm
max,8} = 8.623$ one should realistically consider temperatures in the
range $T_{\rm max,8} = 8.5 - 8.7$. This implies typical ignition radii
in the range 150 - 200 km, though larger values are still possible in
rare cases.

The hottest fluctuations probably begin with (though do not
necessarily end up with) radial velocities lower than the
average. Longer residency in the burning region leads to higher
temperature fluctuations.  Doing the same calculation for an assumed
average radial speed of 20 km s$^{-1}$, one obtains ignition for hot
fluctuations in the range $T_{\rm max,8} = 8.0 - 8.2$ at radii again
near 200 km.  It is important for the issue of multi-point ignition
that the igniting fluctuation takes from one to several seconds to
reach its ignition radius (\Sect{shutoff}).

This maximum fluctuation temperature at runaway, $T_{\rm max}$, is to
be distinguished from the mean central temperature, $T_o$, in that it
reflects conditions in a small, atypical fluid element out on the tail
of the probability distribution function (PDF). Just how far out, is a
critical issue.

\subsection{The probability density function for fluctuations}
\lSect{pdf}

In order to address this question, one needs to assess the probability
that a given high temperature fluctuation will occur. Ignition does
not happen when the average core temperature reaches a flash point,
but when the hottest persistent fluctuation does.  

We assume a simple model that should capture the essence of the real
runaway. Divide the star into two regions: the burning core with mass
$M_{\rm core}$, and a larger reservoir, $M_{\rm conv}$, of cooler
matter to which it is convectively coupled. Because of the
finite speed of convection, $M_{\rm conv}$ shrinks during the
final stages of the runaway while $M_{\rm core}$ remains nearly
constant, but we examine conditions during the last few seconds before
the explosion when this ratio is approximately constant.

Matter passing through M$_{\rm core}$ gets a temperature increment,
$\Delta T$, due to nuclear burning. The magnitude of this increment is
sufficient, when mixed with $M_{\rm conv}$, to heat the entire region
by an amount $\Delta T$ during a convective mixing time. That is, each
time the convection zone circulates once, one must generate an amount
of energy equal to its adiabatic excess, $\Delta \nabla T \, l$, which
is lost.  Since $M_{\rm conv} \gg M_{\rm core}$, the time spent by a
fluid element in the energy generating core, is much less than the
convective mixing time, by approximately the ratio $M_{\rm core}/M_{\rm
conv}$.

An operational value for this ratio comes from comparing the time it
takes matter at the center of the star to appreciably increase its
temperature in the presence or absence of convective cooling. From a
one-dimensional calculation that employs a time-dependent
mixing-length model for convection (the Kepler code, Weaver,
Zimmerman, \& Woosley 1978), for a range of central temperatures $7.0
\, \ltaprx \, T_{o,8} \, \ltaprx \, 7.5$ this ratio is approximately
50.  Very crudely this is also the the ratio of the mass within one
pressure scale height, about 0.3 $\Msun$, to that inside 130 km, the
one-half energy generating region, about 0.01 $\Msun$.  Henceforth we
adopt $M_{\rm conv}/M_{\rm core}$ = 50.  The time that a fluid element
resides in $M_{\rm core}$ is approximately its radius divided by the
convective speed, $\sim 100$ km s$^{-1}$, or 1 second. The total
convective mixing time for $M_{\rm conv}$ is thus $\sim$50 seconds.

During each mixing event in $M_{\rm core}$, a certain fraction of the
material, $f <1$, is exchanged. Given the efficient nature of
convection and the turbulent nature of the core, $f$ may be close
to unity. A representative value $0.9 \, \ltaprx \, f \, \ltaprx \,
0.99$ will be assumed. This implies that each mixing leaves behind
between 1\% and 10\% of the mass in $M_{\rm core}$. This residual
matter increases its temperature beyond the average for the core,
which in fact changes very little. After $n$ mixing events, each
approximately 1 s in duration, the fraction of material that has
increased its temperature by $n \Delta T$ is
$(1-f)^n$. Such a power law naturally gives rise to an exponential PDF
for temperature fluctuations in the core, such that the probability per
unit volume (or mass) for finding a fluid element with temperature, $T$,
when the average is $T_o$, is
\begin{equation}
EPDF \ = \ - \frac{ln (1 - f)}{\Delta T} \, \exp \, \left[ln (1-f)
(T-T_o)/\Delta T\right],
\label{pdfeqn}
\end{equation}
Here, the PDF has been normalized on the temperature domain
($T_o,\infty$). This function gives the probability, per unit
temperature, that a measurement at a random time and place will give a
temperature, $T$. The integral of this function from $T > T_o$ to
infinity is the fraction of the mass that has temperature 
$T$ or greater. Because of the simple toy model assumed here, the PDF is 
exponential (hence, ``E'' in EPDF), but one cannot exclude that 
it might instead be Gaussian. If the fluctuations are statistically 
independent of one another, the central limit theorem implies
\begin{equation}
GPDF \ = \  \left(\frac{4}{\pi (F \Delta T)^2}\right)^{1/2} \, \exp \,
(- \frac{(T - T_o)^2}{(F \Delta T)^2})
\label{gpdfeqn}
\end{equation}
where $F < 1$ is an uncertain factor, not much less than one, to be
determined by experiment or simulation.

Unfortunately, the PDF by itself does not say how the temperature
excess is distributed - in a few large blobs or very many small
ones. For this, one needs additional information or assumptions.
In particular, how many thermally discrete regions exist in the 
inner 100 km where the energy is mostly generated?

\subsection{Persistence of fluctuations}

A minimum size for the thermally discrete cell is capable of crossing
the burning region while retaining its excess heat for a time, $\tau$,
without cooling by conduction is $\lambda_{\rm cond}$, given by
\begin{equation}
\begin{split}
\tau \ &\approx \ \frac{\lambda_{\rm cond}^2}{D} \ = \
\frac{\lambda_{\rm cond}^2 \, \rho \, c_P}{\sigma}\\ &\approx \
\frac{l}{v_{\rm rms}} \ \sim \ 1 {\ \rm s},
\end{split}
\end{equation}
or $\lambda_{\rm cond} \sim 10$ cm. Based upon this scale alone, the
number of thermally discrete regions in $M_{\rm core}$ would be very
large, $\sim 10^{18}$, but turbulence changes this by making the size
of the thermally discrete region much larger..

In a turbulent fluid, large eddies help to generate fluctuations which
the small ones disrupt. Consequently, there is some cut off to the
size spectrum of persistent fluctuations that depends on the intensity
of turbulence in the medium. A relevant length scale is the distance
over which the {\sl average} temperature varies by a characteristic
fluctuation scale. For a nearly adiabatic core,
\begin{equation}
\lambda_{\rm turb} \ \sim \ \frac{\Delta T}{\nabla T_{\rm ad}}.
\end{equation}
For $\Delta T$ given by Eq. \eqref{tfluc} and $\nabla T_{\rm ad}$ by
Eq. \eqref{tadib}, $\lambda_{\rm turb}$ is about a km.

Eddies smaller than this will tend to smooth out fluctuations larger
than $\Delta T$, while larger ones will just move them around and
create new ones. The number of such regions inside 100 km that might
survive against turbulent mixing is about 10$^6$. This is a
more realistic estimate than the larger number given by conduction
alone, and will be used in subsequent discussions. Fortunately the
answer will only depend on the logarithm of this number.

Of course, the turbulence is not really homogeneous or isotropic.
Porter \& Woodward (2000), in three-dimensional numerical studies of
stellar convection up to $Ra \sim 10^{12}$, find that the upwelling
plumes, the ones that would carry the ignition sparks here, are
considerably more laminar than the highly turbulent downdrafts. It is
difficult to quantify this result for the present problem of white
dwarf ignition, but regions of a km or so might survive turbulent
dispersal for the several seconds it takes them to run away.

\subsection{Exponentiation of ignition points}
\lSect{multip}

Once a single point has ignited, will more follow or will the runaway
only ignite once?  When the winner crosses the finish line, how close
behind are the second, third, etc. runners?

We previously derived (\Sect{ign}) $T_{\rm max,8} \approx 8.0 - 8.7$
as the central temperature of the unusually hot spark that finally
ignites the runaway. We shall adopt $T_{\rm max,8} = 8.5$ in what
follows, but similar results are obtained for other values in this
range. At the time of interest, there will be a large number of points
in the core with temperature close to, but less than $T_8 = 8.5$. In
fact the average central temperature at that time will be $T_{o,8} =
8.5 - n \Delta T$ where
\begin{equation}
n \ = \ \frac{log \, N}{|log \, (1-f)|}.
\label{bign}
\end{equation}
For $N = (100 \ {\rm km}/\lambda)^3 \sim 10^6$ and assuming an
exponential PDF with $f = 0.9 - 0.99$, $n = 3 - 6$, which should be
typical. Even if $N \sim 10^{18}$ and $f = 0.9$, $n = 18$, an extreme
upper bound. 

From Eq.\eqref{tfluc} one can evaluate $\Delta T/T$ to determine the range 
of average central temperature when the runaway
finally commences to be
\begin{equation}
\begin{split}
T_{o,8} \ &\approx \ 8.5 \ - \ ({\rm 3 \ to \ 6}) \, \Delta T \\
&\approx \ 7.7 - 7.9.
\end{split}
\end{equation}
The exponentiation time scale, that time during which the number of
ignition points will rise one $e$-fold, is the time it takes for the
temperature near $T_8 = 8.5 - d T$ to rise by $\delta T$, where from 
Eq. \eqref{pdfeqn},
\begin{equation}
\delta T = \Delta T/| ln(1 -f) |.
\label{tfluct1}
\end{equation}
That is, for $f = 0.9$ and an exponential PDF,   
\begin{equation}
\begin{split}
\tau_{\rm exp} \ &= \ \frac{c_{\rm P}}{S_{\rm nuc}} \, \delta T \\
&\approx \ 0.1 \ {\rm s} \
\left(\frac{7.7}{T_{o,8}}\right)^{23} \,
\left(\frac{2}{\rho_{o,9}}\right)^{3.3} \, \left(\frac{\delta
T_8}{0.02}\right).
\label{tauexp}
\end{split}
\end{equation}
A PDF that is Gaussian(Eq. \ref{gpdfeqn}), or a larger value of $f$ would
give a shorter time scale.

For reasonable choices of $f$ and $N$, the answer is considerably less
than the time it takes a spark to traverse the convection region,
about 1 s, and comparable to the time it takes for the supernova
itself to expand and shut off the runaway (\Sect{shutoff}). {\sl Hence
it seems that multi-point ignition is favored, though, within current
uncertainties, not guaranteed.} The number of ignitions per second
will be 
\begin{equation}
\dot n (t) \ = \ \exp(t/\tau_{\rm exp}).
\end{equation}
As noted previously, this answer is relatively insensitive to the
uncertain value of $N$ (Eq. \ref{bign}). Numerical calculations to
confirm this result will need sufficient resolution to follow the
evolution of temperature fluctuations far out on the tail of the PDF,
but are badly needed.

\subsection{The end of ignition}
\lSect{shutoff}

Nuclear burning in the core continues to raise the temperature for a
time, but once ignition has occurred at any point, the spread of the
flame rapidly reduces the binding energy of the white dwarf. Regions
not yet encroached by flame are cooled by this expansion, shutting
off the ignition. From that point onwards, the calculation is purely
one of flame propagation.

When does this occur? We carried out a simulation of a 1.38 $\Msun$
white dwarf at the onset of carbon runaway ($T_{o,8} = 7.6; \rho_{o,9}
= 2.5$) using the Kepler stellar evolution code (Weaver, Zimmerman, \&
Woosley 1978). At that time, the energy input from nuclear burning,
i.e., the convective luminosity had reached $8 \times 10^{45}$ erg
s$^{-1}$, in good agreement with Eq.\eqref{loft}. We then switched off
nuclear energy generation and took a tiny time step, allowing the
model to expand adiabatically with the existing velocity structure.
From this, the rate of adiabatic reduction in central density
appropriate to this luminosity was determined numerically,
\begin{equation}
\frac{d \, ln \, \rho}{d \, t} \ = \ -2.2 \times 10^{-4} \ {\rm s^{-1}}.
\end{equation}
From the equation of state, the adiabatic relations for $d \, ln \,
\rho$ and $d \, ln \, T$ are
\begin{equation}
\begin{split}
\frac{d \, ln \, P}{d \, ln \, \rho} \ &= \ \Gamma_1 \ \approx \
\frac{4}{3} \\ \frac{d \, ln \, T}{d \, ln \, P} \ &= \ \frac{\Gamma_2
-1}{\Gamma_2} \ \approx \ 0.43. \\
\end{split}
\end{equation}
So, expansion in the absence of nuclear burning reduces the 
temperature by 
\begin{equation}
\left(\frac{d \, ln \, T}{d \, t}\right)_{\rm exp} \ = \ -1.2 \times
10^{-4} \ {\rm s^{-1}}.
\end{equation}
Note that this relation between $\delta T$ and $\delta \rho$ is very
different than Eq.\eqref{dp} which is evaluated for non-adiabatic
expansion at constant {\sl pressure}. 

This rate of cooling by expansion should be proportional to the change
in net binding energy, and thus to total rate of energy deposition in
the star. Once the flame forms, this occurs, off center, at a rapidly
increasing rate. Because all motions are initially very subsonic, the
whole star responds, even at its center, to energy deposited anywhere.

In the middle of the star, steady nuclear burning was actually raising
the temperature at a rate that can also be determined numerically, with
the energy generation turned back on,
\begin{equation}
\left(\frac{d \, ln \, T}{d \, t}\right)_{\rm nuc} \ = \ 3.1
\times 10^{-2} \ {\rm s^{-1}}.
\end{equation}
This implies that when the energy generation by the ``flame sheet''
becomes 250 times greater than $L = 8 \times 10^{45}$ erg s$^{-1}$,
i.e., $2 \times 10^{48}$ erg s$^{-1}$, expansion will start to quench
the ignition.

A flame, once born, delivers energy at a rate
\begin{equation}
\dot \epsilon \ = \ q_{\rm nuc} \, A \, u \, \rho 
\end{equation}
where $q_{\rm nuc}$ is the energy released by carbon and oxygen fusing
to iron, about $7 \times 10^{17}$ erg g$^{-1}$, $A$ is its surface
area, and $u$ is the average speed normal to the surface bounded by $A$.
Niemeyer et al (1996), find, for single point off-center ignition, $u
\sim 200$ km s$^{-1}$ after about 0.1 s when the burned region is about
10$^{14}$ cm$^2$ (their Fig. 4), so that $\dot \epsilon \sim 3 \times
10^{48}$ erg s$^{-1}$.

The ignition process thus ceases about 0.1 s after the first spark
ignites off center. Other sparks, however, will still be in transit
since it takes $\sim$ 1 s to emerge from the core (\Sect{ign}). These hot,
localized burning regions, already in the process of running away,
will be more difficult to extinguish. We estimate that the total
ignition process goes on for several tenths of a second.

\subsection{Angular distribution of the sparks}
\label{angular}

Even though the exponentiation time scale in Eq.\eqref{tauexp} is 
short and the number of potential ignition points large, there is
some ambiguity in the counting of discrete points. If the dominant flow
is, for example, dipole in nature (\Sect{flowp}), ignition will occur
preferentially on one side of the star.  The concept of multi-point
ignition becomes blurred if those points are all in close proximity.

Indeed there may be great difficulty getting a viable supernova
explosion if all the ignition occurs on one side (Niemeyer,
Hillebrandt, \& Woosley 1996). Unless a transition to detonation
occurs or pulsational oscillations, it will be difficult to ever burn
the other side. The explosion will then be sub-energetic and produce
too little $^{56}$Ni.

We speculate that the symmetric large scale flow, the dipole term in
the case of spherical convection, is broken at the high Rayleigh
number characteristic of the white dwarf as it runs away. In this
sense the dipole flow in stellar convection is like the large
cell-sized circulation seen in Rayleigh-Benard experiments. The flow
is dominated by the largest symmetric scales. High Rayleigh number as
in Eq. \eqref{rayleigh} may lead to diminished importance of both.
It is also possible that rotation will diminish the effect of the 
dipole (Kuhlen, Woosley, \& Glatzmaier 2002). 

If so, the fluctuations flowing out of the core may be not only
numerous and small, but nearly isotropic. Experiments and numerical
simulations to test this speculation are needed, but will be difficult.

\subsection{Further evolution of sparks}

Our calculations suggest multi-point ignition roughly 150 - 200 km off
center. This is a small amount of mass, at most a few hundredths of a
solar mass. Is the distinction between $r = 0$ and 150 km important?

Yes, it is. Once a spark burns to iron, its density contrast, rather
than being $\Delta \rho/\rho \sim 10^{-4}$ will be 15\%. The radial
acceleration resulting from this buoyancy will be
\begin{equation}
\begin{split}
g_{\rm eff} \ &\approx \ \frac{4 \pi G \rho r}{3} \, \left(\frac{\Delta \rho}{\rho}\right) \\
&\approx \ 8 \times 10^8 \, r_7 \ {\rm cm \ s^{-2}}.
\end{split}
\end{equation}
A blob starting off center accelerates rapidly and does most of its
burning far out in the supernova. In a tenth of a second it is already
moving several hundred km s$^{-1}$. One started dead center only
expands initially at the laminar speed (or at a typical convective
speed, which is comparable) and, in the same 0.1 s, goes only about 10
km.

\section{CONCLUSIONS}

Based upon a simple analytic description of the runaway, we expect
that the first ignition will occur roughly $150 - 200$ km off center
when the hottest point in the center of the star reaches $8.0 - 8.7
\times 10^8$ K (\Sect{ign}). The average central temperature of the
star when these fluctuations are produced is $T_{o,8} \approx 7.7 -
7.9$ (\Sect{multip}). The orientation of this point with respect to
the star's center may be given by the existence of a strong dipole
asymmetry to the macroscopic convective flow just prior to the runaway
(\Sect{flowp}).

The existence of this dipole component (\Sect{flowp}), seen previously
in numerical models of red-giant convection, in main sequence massive
stars, and further explored in Paper II, is important in determining
the outcome of the explosion. Even given multiple ignition points,
if they all happen on the same side of the star in close proximity,
and if there is no subsequent detonation, a weak explosion will
ensue. It may be, however, that quasi-symmetric flow is restored at
higher values of $Ra$ than have been simulated thus far. By analogy to
Rayleigh-Benard convection, such a restoration of symmetry might be
expected in the Kraichnan regime (\Sect{flowp}). Alternatively,
rotation may be invoked to alter the dipole flow (Paper II).

The convective speed, which gives a measure of the turbulent energy
input on large length scales, is 50 to 100 km s$^{-1}$ throughout most
of the mass of the white dwarf when it explodes. This, not the laminar
flame speed gives a minimum rate at which the flame spreads, even in
the absence of strong Rayleigh-Taylor instability.

To carry the stellar luminosity at this speed, a superadiabatic
excess, $\Delta T$, is required across the burning
region. Mixing-length theory and Nusselt-Rayleigh scaling give a
common prediction for the characteristic super-adiabatic excess,
$\Delta T$, in the convection zone, i.e., $Ra = (Nu Pr)^{1/2}$, only
in the Kraichnan regime. All other scalings give a value for $\Delta
T$ that depends upon the conductivity or viscosity or both. Assuming
the validity of $(NuPr)^{1/2}$ scaling, and based upon a simple toy
model (\Sect{pdf}), we show that the distribution of temperature
fluctuations in the burning core may be exponential with a
characteristic temperature scale $\ltaprx \Delta T$. The hottest
fluctuations are not produced by compression as in H\"oflich \& Stein
(2002), but by extended residency and burning near the center of the
star. We give an expression for the expected PDF, Eq.\eqref{pdfeqn}.

Once the first point has ignited, the delay until the second, third,
etc. ignites is given by the time it takes a fluid element on the tail
of the PDF to raise its temperature by an amount $\Delta T/| \, ln \,
(1-f) |$ where $f$ is the fraction of the core's mass that gets
exchanged during a mixing time scale. For reasonable assumptions, this
delay is $\tau_{ign} \sim 0.1$ s, Eq.\eqref{tauexp}. This is small
compared with the time it takes the first igniting spark to transverse
the burning region ($\sim 1$ s; \Sect{ign}) and comparable to the time
for the expansion of the supernova to shut off the ignition process
(\Sect{shutoff}). The number of ignition events
increases as approximately $\exp (t/\tau_{ign})$. Thus multi-point
ignition is possible.

However, an important unresolved issue is the angular distribution and
mean separation of ignition points. This will be influenced by the
dipole flow, but also by the propensity of hot regions to cluster.

Given that the number of ignitions and their location is exponentially
sensitive, by Eq.\eqref{tauexp}, to small variations in the conditions
in the core at the time of runaway, one expects that SN Ia models,
starting from nearly identical circumstances will display chaotic
variation in their properties, including their light curves. Factors
that might cause a systematic variation in this dispersion include the
carbon-mass fraction (Eqs.\ref{snuc} and \ref{vlam}) and the
central density at ignition. Calculations to quantify these
dependences will be difficult, but should some day be undertaken given
the importance of Type Ia supernovae as standard candles for
cosmology.

Our analytic study can also serve to motivate and guide future
numerical work. First, because the issue of the dipole nature of the
flow is important, all calculations should carry the entire $4 \pi$
solid angle of a sphere. Calculations that study just a wedge and
assume spherical symmetry will miss an important aspect of the
problem. Because of concerns that the dipole might be artificially
created by the tendency of turbulence to cascade to large scales in 2D
simulations, some calculations, at least, will need to be done in 3D.

A major goal of numerical simulation should be to ascertain the PDF
for the temperature fluctuations. Is it exponential, as assumed here,
or Gaussian? Is the characteristic scale given by $\Delta T$ in 
Eq.\eqref{tfluct1}? What is the operational value of $f$? What is the
angular distribution and size distribution of the fluctuations?  The
Rayleigh number of such studies needs to be as high as possible, first
because $Ra$ in the star is naturally unapproachably high, and second,
to see the dependencies of all the above answers, including the dipole
flow, on $Ra$. In Paper II we shall address some of these questions.

This work has been supported by the NSF (AST 02-06111), NASA
(NAG5-12036), and the DOE Program for Scientific Discovery through
Advanced Computing (SciDAC) under grant DE-FC02-01ER41176.  We greatly
appreciate interesting discussions with Alan Kerstein, Jens Niemeyer,
and Alex Heger that helped to clarify some of the ideas presented here.

\end{document}